\documentclass[aps,
               pra,
               twocolumn,
               nofootinbib,
               floatfix,
               showpacs
              ]{revtex4}

\usepackage{amsmath, amssymb}
\usepackage{graphics}
\usepackage{pstricks}
\usepackage{graphicx}
\usepackage{color}

\renewcommand{\i}{\mathrm{i}}

\renewcommand{\vector}[1]{\mathbf{#1}}

\newcommand{\cred}[1]{{\color{red}{#1}}}

\newcommand{\weg}[1]{{\cred{#1}}}

\renewcommand{\weg}[1]{}

\newcommand{\del}{\partial}

\newcommand{\Eq}[1]{Eq.~(\ref{eq:#1})}

\newcommand{\Fig}[1]{Fig.~\ref{fig:#1}}
\newcommand{\fig}[1]{\ref{fig:#1}}

\begin{document}

\title{Superfluid Turbulence: Nonthermal Fixed Point in an Ultracold Bose Gas}

\author{Boris Nowak}
\author{D\'enes Sexty}
\author{Thomas~Gasenzer}
\email{t.gasenzer@uni-heidelberg.de}
\affiliation{Institut f\"ur Theoretische Physik,
             Ruprecht-Karls-Universit\"at Heidelberg,
             Philosophenweg~16,
             69120~Heidelberg, Germany}
\affiliation{ExtreMe Matter Institute EMMI,
             GSI Helmholtzzentrum f\"ur Schwerionenforschung GmbH, 
             Planckstra\ss e~1, 
             64291~Darmstadt, Germany} 

\date{\today}

\begin{abstract}
Nonthermal fixed points of far-from-equilibrium dynamics of a dilute degenerate Bose gas are analysed in two and three spatial dimensions.  
For such systems, universal power-law distributions, previously found within a nonperturbative quantum-field theoretic approach, are 
shown to be related to vortical dynamics and superfluid turbulence.
The results imply an interpretation of the momentum scaling at the nonthermal fixed points in terms of independent vortex excitations of the superfluid.
Long-wavelength acoustic excitations on the top of these are found to follow a non-thermal power law.
The results shed light on fundamental aspects of superfluid turbulence and have strong potential implications for related phenomena studied, e.g., in early-universe inflation or quark-gluon plasma dynamics.
\end{abstract}


\maketitle


From the formation of Bose-Einstein condensates in ultracold gases to quark-gluon plasmas produced in heavy-ion collisions, over a range of twentyfour orders of magnitude in temperature, nonequilibrium dynamics governs many interesting phenomena. 
Turbulence is an outstanding and intricate example, which can be described as an oriented stationary flow of energy or particles between different scales. 
This idea of an energy cascade goes back to the work of Richardson in the context of atmospheric science \cite{Richardson1920a}.
Kolmogorov, in his 1941 mathematical discussion of turbulence in an incompressible fluid, added the concept of universality and scaling \cite{Kolmogorov1941a}.
More recently, turbulence has been studied in the context of the inflationary early universe as well as of strongly correlated matter produced in heavy-ion collisions \cite{Micha:2002ey,Berges:2008wm,Berges:2008sr,Scheppach:2009wu,Arnold:2005ef}.
Dynamical critical phenomena, i.e., fixed points of the evolution away from thermal equilibrium have been proposed.
They potentially affect the equilibration process by forcing the evolution to critically slow down before final thermalization.
New scaling laws were found by analysing non-perturbative Kadanoff-Baym dynamic equations  \cite{Berges:2008wm,Berges:2008sr}.
Analogous predictions for a nonrelativistic Bose gas were given in \cite{Scheppach:2009wu}, proposing strong matter-wave turbulence in the regime of long-range excitations.

Superfluid turbulence, also referred to as quantum turbulence (QT) has been the subject of extensive studies in the context of helium \cite{Halperin2008a}. 
In contrast to eddies in classical fluids vorticity in a superfluid is quantized \cite{Onsager1949a,Feynman1955a}, and the creation and annihilation processes of quantized vortices are distinctly different \cite{Halperin2008a}. 
The observation of a Kolmogorov 5/3-law \cite{Kolmogorov1941a} in experiments with superfluid helium, cf.~\cite{Vinen2002a} for a review, received much attention 
\cite{Nore1997a,Araki2002a,Kobayashi2005a,Horng2007a,Tsubota2010a,Yukalov2010,Davis2010}. 
In particular, the role of the normal-fluid as compared to the superfluid component in the turbulent flow is under debate \cite{Halperin2008a}.
In the context of the kinetics of condensation and the development of long-range order in a dilute Bose gas, the role of turbulence in the superfluid and its acoustic excitations was discussed in Refs.~\cite{Kagan1992a,Berloff2002a}, see also \cite{Kozik2009a} for a recent review.
A possible observation of QT in ultracold atomic gases presently poses an exciting task for experiments \cite{Anderson2008,Henn2009a}. 

A satisfactory ab-initio mathematical description of both quantum and classical turbulence is inherently difficult due to the generically strong correlations building up within the system. 
Analytical results are known, however, in regimes where kinetic theory applies:
In a dilute, degenerate Bose gas the normal-fluid component can vary at the expense or gain of the superfluid part. 
As a consequence, the gas is compressible and so-called weak wave-turbulence phenomena can occur. 
For these, scaling laws can be derived by analysing kinetic equations \cite{Zakharov1992a}. 

Recent developments presented in Refs.~\cite{Berges:2008wm,Berges:2008sr,Scheppach:2009wu,Carrington:2010sz} allow to set up a unifying description of scaling, both in the ultraviolet quantum-Boltzmann kinetic regime and in the infrared (IR) regime of long wavelengths where perturbative approximations break down. 
In the IR regime, new scaling laws were found \cite{Berges:2008wm} by analysing non-perturbative Kadanoff-Baym dynamic equations with respect to nonthermal stationary solutions. 

Here, we study QT in two and three dimensional dilute Bose gases by means of simulations in the classical-wave limit of the underlying quantum field theory. 
We compare the scaling exponents of the single-particle momentum distribution and their analytical predictions~\cite{Scheppach:2009wu}.
While we find excellent agreement, our results provide an interpretation of the dynamical fixed points proposed in Refs.~\cite{Berges:2008wm,Scheppach:2009wu} for the case of an ultracold Bose gas: 
The appearance of nonperturbative infrared scaling reflects the presence of independent vortices.
Moreover, the power spectra of underlying compressible excitations suggest an understanding in terms of acoustic turbulence \cite{Zakharov1992a} on the top of the vorticity-bearing quasicondensate.\\


\textit{Scaling laws: Analytical Predictions.}~Based on the concept of universality, the turbulent single-particle spectrum $n(\vector{k})$ is assumed to scale as
\begin{equation}
n(s\vector{k}) = s^{-\zeta}n(\vector{k}),
\end{equation}
where $s$ is some positive real
number. This scaling law is compatible with a more general scaling behaviour assumed for the full 
two-point correlation function of the bosonic quantum field as discussed in Ref.~\cite{Scheppach:2009wu}.

To determine the positive exponent $\zeta$ in the infrared, where $n(\vector{k})\sim|\vector{k}|^{-\zeta}$ is large, an approach beyond kinetic weak-wave-turbulence theory has been used. 
This is based on analysing fixed points of Kadanoff-Baym dynamic equations derived from the two-particle irreducible (2PI)
effective action or $\Phi$-functional \cite{Luttinger1960a}.
While in the regime of large wave numbers the approach goes over into the kinetic description of weak wave turbulence, an effectively renormalised many-body $T$-matrix modifies the scaling law in the IR.
This is a consequence of the resummation of a particular class of Feynman diagrams contributing to the 2PI effective action \cite{Berges:2008wm,Scheppach:2009wu}, a formulation substantially beyond the low-order perturbative expansion used for the kinetic quantum Boltzmann equation.  
Physically, the renormalised $T$-matrix implies a reduction of the effective interaction strength in the IR regime of strongly occupied modes \cite{Berges:2008wm}. 
As a consequence, single-particle occupation
numbers rise, towards smaller wave numbers, in a steeper way than in the weak-turbulence regime.
The IR scaling exponent for constant radial quasiparticle flow in $d$ dimensions was predicted in Ref.~\cite{Scheppach:2009wu} as
\begin{align}
  \zeta = \zeta^\mathrm{IR}_{\mathrm{Q}}
  &= d+2 .
  \label{eq:kappaIRQ}
\end{align}
%

%
\textit{Semiclassical field simulations.}~A dilute superfluid Bose gas can be described, in the classical-wave limit, by the Gross-Pitaevskii equation (GPE)
\begin{equation}  \label{GPE}
   \i \del_t \psi(\mathbf{x},t)= \left[ -\frac{\nabla^2}{2m}+g|\psi(\mathbf{x},t)|^2 \right] \psi(\mathbf{x},t) .
\end{equation}
(In our units $\hbar=1$). 
$m$ is the boson mass, and $g$ quantifies the interaction strength. We consider a gas in a box of size $L^d, d=2,3$, with periodic boundary conditions. 
Diluteness implies that $a\ll l$, the interparticle spacing $l=\overline{n}^{-1/d}$ being determined by the mean density $\overline{n}=N/L^d$. 
The initial values for the real and imaginary parts of the field $\psi(\mathbf{k},0)$ were randomly chosen from a Gaussian  distribution with width $1/2$, centred around $\sqrt{n(\mathbf{k},0)}\exp\{i\phi(\mathbf{k},0)\}$, where $n(\mathbf{k},t)=\langle \psi^\dag(\mathbf{k},t)\psi(\mathbf{k},t)\rangle$ is the occupation number at time $t$ and $\phi(\mathbf{k},0)$ is a random phase angle. Correlation functions, including $n(\mathbf{k},t)$, were obtained by averaging over many trajectories. We have simulated the time evolution on a grid with side length $L=N_s a_s$,  lattice spacing $a_s$ in $d=2$, $3$. To induce transport from small to large wave numbers, only a few modes near $\mathbf{k}=0$ were chosen to be occupied at the initial time with $n(\mathbf{k},0)\gg1$. Such an initial state can be prepared, e.g., by Bragg scattering of photons from a Bose-Einstein condensate. 
During the evolution to a new equilibrium state, energy is transported to modes with higher wave number, and, given appropriate interaction strength, one observes quasistationary momentum distributions to develop prior to final thermalisation.

During the intermediate stage of the equilibration quantised vortices appear. 
To exhibit the vortical flow we use the Madelung representation $\psi(\mathbf{x},t)=\sqrt{n(\mathbf{x},t)}\exp\{i\phi(\mathbf{x},t)\}$ of the field in terms of the density $n(\mathbf{x},t)$ and a phase angle $\phi(\mathbf{x},t)$. 
This allows to express the particle current $\mathbf{j}=i(\psi^*\nabla \psi - \psi \nabla \psi^*)/2=n\mathbf{v}$ in terms of the velocity field $\mathbf{v}=\nabla\phi$.
\begin{figure}[!t]
 \includegraphics[width=0.47\textwidth]{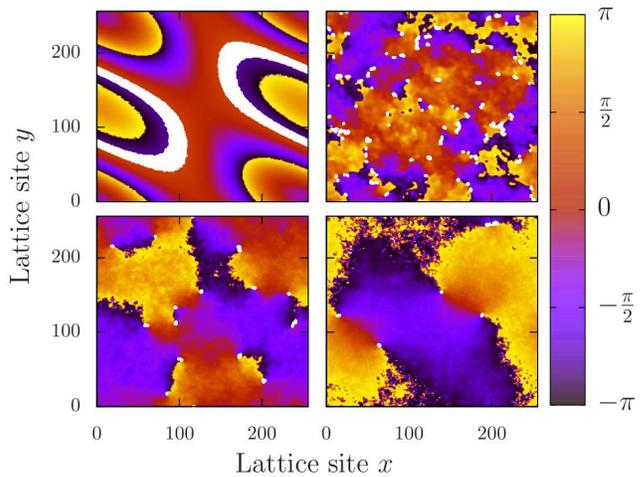}
 \caption{Phase angle $\phi(\vector{x},t)$ (color scale) at four times during a single run of the simulations in $d=2$. 
The white spots mark vortex cores where the density falls below 5\% of the mean density $\overline{n}$. 
Parameters are: 
$\overline{g}=2mga^{2-d}_s=3\cdot 10^{-5}$, $N=10^8$, $N_s=256$, $\overline{t}= t/(2ma^2_s)$. 
Shown times are: 
1. $\overline{t}=26$ (top left): Ordered phase shortly after initial preparation. 
2. $\overline{t}=820$ (top right): After creation of vortex-antivortex pairs. 
3. $\overline{t}=6550$ (bottom left): During critical slowing down of the vortex-antivortex annihilation. 
4.  $\overline{t}=10^5$ (bottom right): Low-density vortex-antivortex pairs before final thermalisation.}
\label{fig:PhaseEvolution}
\end{figure}
 \begin{figure}[!t]
 \includegraphics[width=0.36\textwidth]{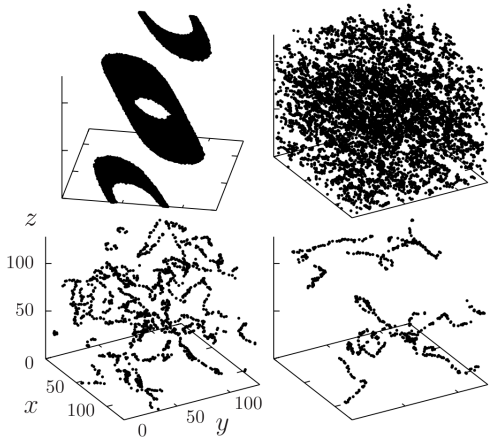}
 \caption{Snapshots of a single run of the evolution in $d=3$ dimensions. The points show where the density falls below 5\% of the mean density $\overline{n}$. 
 Parameters are:  $\overline{g}=8\cdot 10^{-4}$ , $N=10^9$, $N_s=128$. 
1. $\overline{t}=0$ (top left) 
 2. $\overline{t}=103$ (top right)
 3. $\overline{t}=410$ (bottom left)
 4. $\overline{t}=1640$ (bottom right)}
 \label{fig:3dPhaseEvolution}
 \end{figure}
In \Fig{PhaseEvolution} we show the phase angle $\phi(\mathbf{x},t)$ during a single run in $d=2$ dimensions,  while in \Fig{3dPhaseEvolution} the evolution in $d=3$ is illustrated by the near-zero-density points.
Many singly quantized vortex-antivortex pairs (vortex rings) are formed once the initial coherent wave has developed to form shock-wave-like fronts. 
During the later evolution, the pairs mutually annihilate at a substantially decreasing rate. After the last pair has vanished the gas thermalises. 
In $d=3$, decreasingly tangled vortex lines are observed during the near-stationary evolution.
The initial conditions lead to a dense chaotic tangle, as studied also in Refs.~\cite{Berloff2002a,Kozik2009a}. 

\begin{figure}[!t]
\includegraphics[width=0.48\textwidth]{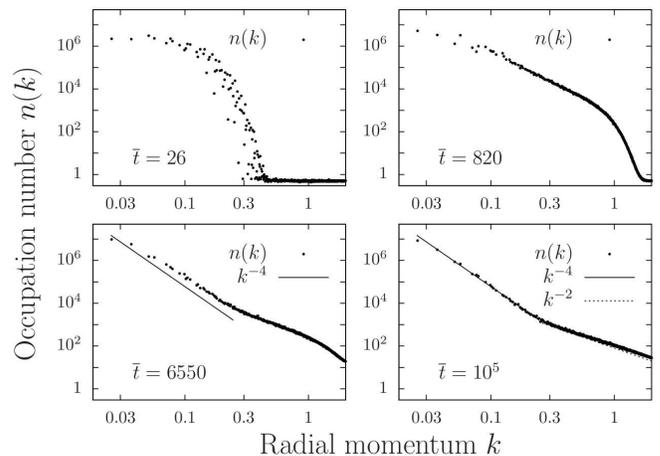}
\caption{Single-particle mode occupation numbers as functions of the radial momentum $k=[\sum_{i=1}^d 4\mathrm{sin}^2 (k_i/2)]^{1/2}, \mathbf{k}=2\pi\mathbf{n}/N_s, \, \mathbf{n}=(n_1,...,n_d),\,  n_i= -N_s/2,...,N_s/2$, for the four different times of the run in $d=2$ dimensions shown in \Fig{PhaseEvolution}.
Note the double-logarithmic scale. An early development of a scaling $n(k)\sim k^{-4}$ is followed by a bimodal scaling with $n(k)\sim k^{-2}$ at larger wave numbers.
}
\label{fig:2dSpectrumEvolution}
\end{figure}

\begin{figure}[!t]
\includegraphics[width=0.45\textwidth]{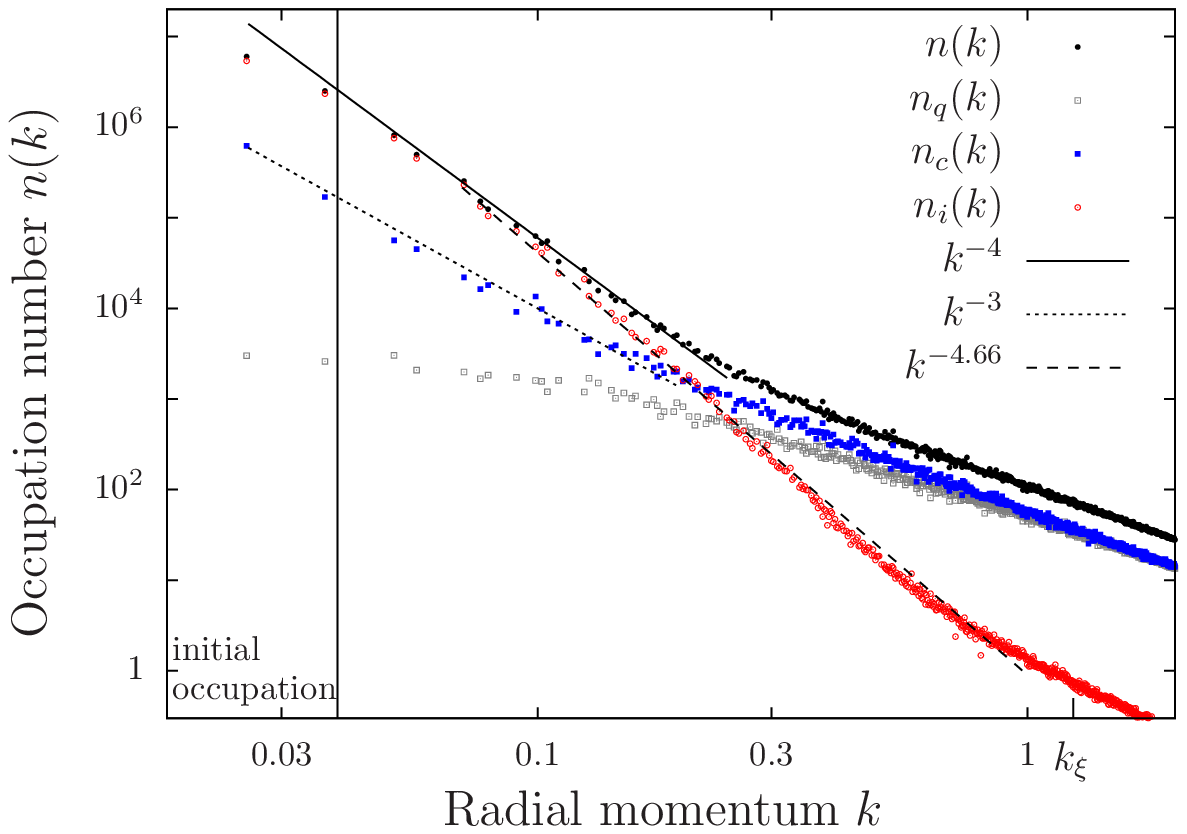}
\caption{Single-particle occupation numbers of different fractions of the system, at time $\overline{t}=10^5$ of the $d=2$ run shown in \Fig{2dSpectrumEvolution}.
The black points are the same as those shown in the lower right panel of \Fig{2dSpectrumEvolution}.
Colors distinguish the fractions $n_\mathrm{i}$ (red circles), i.e., the divergence-free part of the flow field $\vector{w}_{\mathrm{v}}=\sqrt{n}\vector{v}$, $n_{\mathrm{c}}$ (filled blue squares), i.e., the solenoidal part of $\vector{w}_{\mathrm{v}}$, and the quantum-pressure part$n_{\mathrm{q}}$ (open grey squares).
A scaling with $k^{-4.66}$ corresponds to a $5/3$ law for the kinetic energy in $d=2$.
$k_{\xi}=2\pi/\xi$ marks the scale corresponding to the healing length $\xi=1/\sqrt{2mgn}$}
\label{fig:2dSpectrumComparison}
\end{figure}

\begin{figure}[!t]
\includegraphics[width=0.45\textwidth]{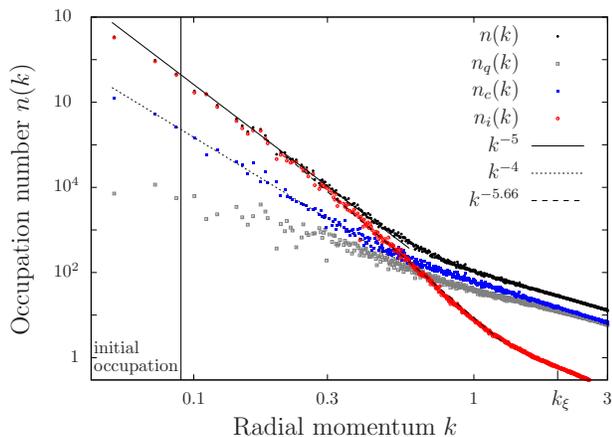}
\caption{Same as in \Fig{2dSpectrumComparison}, for the run in $d=3$ dimensions shown in \Fig{3dPhaseEvolution}, at the time $\overline{t}=1640$ of the situation in its lower right panel. 
A scaling with $k^{-5.66}$ corresponds to a $5/3$ law for the kinetic energy in $d=3$.}
\label{fig:3dSpectrumComparison}
\end{figure}

Power-law behaviour is found in the momentum distributions of particles and energy as we discuss in the following.
\Fig{2dSpectrumEvolution} shows the distribution of occupation numbers $n(k)=\int \mathrm{d}\Omega_d \, n(\mathbf{k})$ over the radial momentum $k=|\mathbf{k}|$, for the four snapshots of the two-dimensional system depicted in \Fig{PhaseEvolution}, on a double-logarithmic scale. 
($\mathrm{d}\Omega_d$ denotes the angular differential.) 
During the initial evolution the mode occupations gradually spread to larger wave numbers. 
As soon as vortex-antivortex pairs appear, a power-law regime $n(k)=k^{-\zeta}$ is observed. 
During the final stage of the vortex-bearing phase two distinct power laws develop which are found to be in excellent agreement with the analytical prediction of Ref.~\cite{Scheppach:2009wu}. 
While in the ultraviolet the exponent $\zeta^{\mathrm{UV}} = d=2$ exhibits weak wave turbulence, in the infrared the exponent confirms the result $\zeta=d+2=4$, see \Eq{kappaIRQ}, corroborating results for a relativistic model in \cite{Berges:2008wm}. 
Note that in $d=2$, the weak-turbulence exponent $\zeta^{\mathrm{UV}}=2$ is identical to that in thermal equilibrium in the Rayleigh-Jeans regime, $n(k)\sim T/\omega \sim T/k^2$ \cite{Zakharov1992a}.
%
%
The momentum distributions during this final phase are shown again, for $d=2$ and $d=3$, in Figs.~\fig{2dSpectrumComparison} and \fig{3dSpectrumComparison} (filled black circles), respectively, confirming the analytical prediction of  \Eq{kappaIRQ} also for $d=3$.
The scaling $n(k)\sim k^{-2}$ in the ultraviolet reflects that the corresponding modes are already thermalised.
We remark that, for weaker interaction strengths, the weak-wave-turbulence scaling with $\zeta^{\mathrm{UV}} = d=3$ is also seen, at intermediate times.

To understand these findings in the context of QT we analyse kinetic-energy spectra as in \cite{Nore1997a}.
The total kinetic energy $E_{\mathrm{kin}}= \int \mathrm{d}^dx \, \langle |\nabla \psi(\mathbf{x},t)|^2\rangle/2$ can be split, $E_{\mathrm{kin}} = E_{\mathrm{v}} + E_\mathrm{q}$, into a `classical' part $E_\mathrm{v}= \int \mathrm{d}^dx \, \langle |\sqrt{n}\mathbf{v}|^2 \rangle/2 $ and a `quantum-pressure' component $E_\mathrm{q}=\int \mathrm{d}^dx \, \langle |\nabla \sqrt{n}|^2 \rangle/2 $.
The radial energy spectra for these fractions involve the Fourier transform of the generalised velocities $\vector{w}_{\mathrm{v}}=\sqrt{n}\vector{v}$ and $\vector{w}_{\mathrm{q}}=\nabla\sqrt{n}$,
\begin{eqnarray}
 E_{\delta}(k)= \frac{1}{2} \int k^{d-1} \mathrm{d}\Omega_d \, \langle |\mathbf{w}_{\delta}(\mathbf{k})|^2 \rangle,\quad \delta=\mathrm{v},\mathrm{q}.
\end{eqnarray}
Note that the superfluid velocity $\mathbf{v}=\nabla\phi$ of a single vortex diverges as $1/r$ with the distance $r$ from the vortex core.
Hence, the Fourier transform of $\mathbf{v}$ is ill-defined and requires regularisation of $|\mathbf{v}(\mathbf{x})|$ in the vicinity of the core.
In $\mathbf{w}_{\mathrm{v}}$, this is naturally achieved by the factor $\sqrt{n}$ which rises linearly in $r$ for small $r$. 
Following Ref.~\cite{Nore1997a} the regularised velocity $\mathbf{w}_{\mathrm{v}}$ is furthermore decomposed into `incompressible' (divergence free) and `compressible' (solenoidal) parts, $\mathbf{w}_{\mathrm{v}}=\mathbf{w}_{\mathrm{i}}+\mathbf{w}_{\mathrm{c}}$, with $\nabla \cdot \mathbf{w}_\mathrm{i}=0$, $\nabla \times \mathbf{w}_\mathrm{c}=0$, to distinguish vortical superfluid and rotationless motion of the fluid.

For comparison of the kinetic-energy spectrum with the single-particle spectra $n(k)$, we determine occupation numbers corresponding to the different energy fractions as $n_\mathrm{\delta}(k) =  k^{-d-1}E_{\mathrm{\delta}}(k)$, $\delta\in\{$i, c, q$\}$.
In the cases we consider, the resulting spectra $n_\mathrm{i}(k)$, $n_\mathrm{c}(k)$, and $n_\mathrm{q}(k)$ add up to the single-particle spectrum $n(k)$ discussed before. The components and their sum are shown seperately, for $d=2$ and $d=3$, in Figs.~\fig{2dSpectrumComparison} and \fig{3dSpectrumComparison}, respectively.
Red circles denote $n_{\mathrm{i}}$, filled blue squares show the dependence of $n_{\mathrm{c}}$, and open grey squares that of $n_{\mathrm{q}}$. 
Qualitatively, the results are the same for $d=2$ and $d=3$.  
Excitations with large wave numbers are thermally distributed.
In this regime, the spectrum $n(k)\sim k^{-2}$ is dominated by the compressible and quantum-pressure components. 
For smaller wave numbers the scaling changes to $n(k)\sim k^{-d-2}$, being dominated by the velocity $\mathbf{w}_{\mathrm{v}}$.
Moreover, we find that it is this decomposition into $n_{\mathrm{i}}$ and $n_{\mathrm{c}}$ which, for intermediate wave numbers, allows the incompressible part of the energy to develop a Kolmogorov-like scaling $\sim k^{-5/3-d-1}$ above the scale $k_{l}\sim 2\pi/l$ determined by the mean distance $l$ between vortex cores.
While the scaling of the sum of these components is predicted by the field-theoretic analysis in Ref.~\cite{Scheppach:2009wu}, a rising compressible part allows the incompressible contribution to deviate from the IR power law and to develop the observed scaling.
Towards the IR limit, the compressible part becomes too weak such that the scaling of $n_{\mathrm{i}}$ goes over to $\zeta=d+2$.

One can show that the analytically predicted infrared power law $n(k)\sim k^{-4}$ in $d=2$ is consistent with a finite density of independent vortices and antivortices.
The IR scaling $\sim k^{-d-1}$ of the compressible (blue) component suggests an interpretation in terms of acoustic turbulence \cite{Zakharov1992a,Kagan1992a,Kozik2009a}.
Our results show that this component survives for a limited period beyond the time when all vortical excitations have mutually annihilated.
The observed scaling corroborates the numerical findings of Ref.~\cite{Khlebnikov2002a}, whereby we refrain from sharing the interpretation of the power law.

In summary, our results show a distinct power-law behaviour $k^{-\zeta}$ of the single-particle momentum spectrum $n(k)$ as well as of different components of the kinetic-energy distribution over the radial wave number $k$.
Scaling exponents $\zeta$ of $n(k)$ corroborate the analytical predictions of Ref.~\cite{Scheppach:2009wu}. 
Our findings suggest that local field expectation values and short- to intermediate-range coherence, including topological excitations,  are at the basis of the infrared power laws predicted within nonperturbative dynamical field theory \cite{Berges:2008wm,Berges:2008sr,Scheppach:2009wu,Carrington:2010sz}.
For the chosen generic initial conditions, excitations on the top of this classical field background support an interpretation in terms of acoustic wave turbulence.
The connection of these phenomena with ab-initio dynamical field theoretic methods points a way to unified analytical studies of turbulence.
Moreover, it provides hints of how the proposed dynamical critical phenomena  in relativistic systems \cite{Berges:2008wm,Berges:2008sr,Carrington:2010sz} can be realized in nature.

Experimental studies of universal phenomena in nonequilibrium dynamics of ultracold atoms have great potential since universal effects
do not depend significantly on initial conditions and details of the system.   
The study of turbulence in ultracold gases may have great impact on many other fields of physics. 
Prominent examples are strongly correlated nuclear matter produced in heavy-ion collisions and early-universe cosmology.

\textit{Acknowledgements}.
T.~G. and D.~S. thank J. Berges and C. Scheppach for collaboration on related work. 
The authors thank C. Bodet, E. Calzetta, B. Eckhardt, L. Glazman, C. Gross, H. Horner, M. Kronenwett, M.~K. Oberthaler, J.~M. Pawlowski, M. Schmidt, J. Schole, and B. Svistunov for useful discussions. 
They acknowledge support by the Deutsche Forschungsgemeinschaft, by the University of Heidelberg (FRONTIER), and by the Helmholtz Association (HA216/EMMI).


\end{document}